\documentclass[showpacs,amsmath,nofootinbib,amssymb,floatfix, preprint]{revtex4}
\usepackage{graphicx}
\newcommand{\mathsym}[1]{{}}

\begin{document}
\begin{abstract}
We propose a general technique to solve the classical many-body problem with radiative damping. We modify the short-distance structure
of Maxwell electrodynamics. This allows us to avoid runaway solutions \textit{as if} we had a covariant model of extended particles. The resulting equations of motion 
are functional differential equations (FDEs) rather than ordinary 
differential equations. 
Using recently developed numerical techniques for stiff FDEs, we solve these equations for the one-body central force problem with radiative damping with a view to benchmark our new approach.
Our results indicate that locally the 
magnitude of radiation damping may be well approximated by 
the standard third-order expression but the global properties of our solutions
are dramatically different. We comment on the two body
problem and applications to quantum field theory and quantum mechanics.
\end{abstract}
\title{Radiative Damping and Functional Differential Equations}
\author{C. K. Raju}
\affiliation{Indian Institute of Advanced Study, Rashtrapati Nivas, Shimla, 171 005, India}
\altaffiliation[Permanent Address: ]{Inmantec, National Highway 24, Dasna Crossing, Ghaziabad, 201 009, Delhi NCR, India}
\email{c.k.raju@inmantec.edu}
\author{Suvrat Raju}
\affiliation{Department of Physics, Harvard University, Cambridge, MA 02138, USA}
\email{suvrat@post.harvard.edu}
\maketitle

\section{Introduction}
\subsection{Motivation}
It is remarkable how little is known about classical solutions of relativistic many-body electrodynamics with radiative damping. Textbooks often describe simple non-relativistic solutions to the one-body problem with an external force and radiative damping. Studies in accelerators and antennas also consider radiation emitted from charged particles. However, when many bodies are included and the back-reaction of the radiation on the particles involved is considered, the resulting equations are amenable neither to analytic solutions {\it nor} to numerics. We explain the reason for this in greater detail below but
we emphasize here that
the difficulties involved are not merely technical but relate to important questions about the nature of classical electrodynamics.
 
In fact, until recently, there has been {\it no} systematic development of techniques to integrate the equations of relativistic many-body electrodynamics with radiative damping. One reason for this is that, hitherto, there has been no compelling physical reason to consider the intricacies of this problem. For example, in astrophysical processes, the radiation emitted by charged bodies is of interest; however the {\it back-reaction} of this radiation on the bodies is small compared to other forces. Hence, the radiation reaction is often ignored or included only heuristically. 

In recent years, this situation has changed. Simulations of gamma ray bursts, for example, have reached a level of sophistication that authors have started to investigate the influence of radiation reaction \cite{noguchi2004rep}. A more immediate motivation for this study comes from papers of \cite{raju2004ebp,deluca1998ehr,deluca2006stf}. In these papers, it was conjectured that some classical orbits of the many-body system have special properties when radiation reaction is considered. However, further progress on this front was prevented by the absence of techniques to integrate the many-body system\footnote{We explain the reason for our emphasis on {\it many-body} in Section \ref{onevsmany} below} with radiation damping.

In this paper, we develop such a technique. Our technique may be applied to check the conjectures of \cite{raju2004ebp, deluca1998ehr,deluca2006stf}.  Moreover, this technique is of independent interest because it is applicable to the investigation of radiative damping effects in astrophysical processes. It would also be interesting to include the effect of radiative damping in molecular-dynamic simulations carried out in  codes such as {\sc charmm \cite{brooks1983sds,mackerelljr1998cef}, amber\cite{pearlman1995apc}} or {\sc  gromos\cite{scott1999gbs}}.  

In this paper, we elaborate our technique and explain its advantages. Furthermore, we {\it bench-mark} our technique by applying it to the one-body problem. This is feasible since, as we explain further in section \ref{onevsmany}, some solutions to the {\it one-body} Lorentz-Dirac equation have already been obtained. 
Our technique agrees, excellently, with these solutions. Having established the validity of our methods, we hope to apply them to the concrete physical problems described above in a subsequent paper.

\subsection{Setting}
Consider a point particle with mass $m$, charge $q$ and world-line $\alpha^{\mu}(\tau)$ parameterized by its proper time $\tau$. In the presence of an external electromagnetic field strength
 $F_{\mu \nu}^{\rm ext}$ the equation of motion of this particle is given by:
\begin{equation}
\label{eqmotionparticle}
m {\ddot{\bf \alpha}}_{\mu} = q \dot{\alpha}^{\nu} F_{\mu \nu}^{\rm ext}({\bf \alpha}_{\rho}(\tau))  + {q^2 \over 6 \pi \epsilon_0 c^3} \left[\dddot{{\bf \alpha}}_{\mu} - \dot{{\bf \alpha}}_{\mu} {\ddot{{\bf \alpha}}^{\nu}\ddot{{\bf \alpha}}_{\nu} \over c^2} \right].
\end{equation}
The second term on the right may be understood as the regularized action of the particle on {\it itself}. We emphasize that the structure of this term is quite delicate and it {\it cannot} be modified without modifying the Maxwell theory itself. 

This is a third order differential equation and requires 3 initial conditions. However, for a generic choice of initial conditions $\alpha(0), \dot{\alpha}(0), \ddot{\alpha}(0)$, we obtain a runaway solution where the velocity of the particle increases continuously in the future! This solution is clearly unphysical and the existence of such runaways is an embarrassment for the classical theory of electromagnetism. Several solutions to this problem have been proposed in the literature. We discuss some of these,
in Section \ref{otherapproachesubsection} and Appendix \ref{othermethods}. However, one common solution to this problem is as follows. Let us say, we assign the initial acceleration to satisfy
\begin{equation}
\label{plasstransform}
\ddot{\alpha}_{\mu} (0) = {1 \over \kappa} \int_0^{\infty} d \tau e^{-\tau \over \kappa} \left[q \dot{\alpha}^{\nu} F^{\rm ext}_{\mu \nu} ( \alpha_{\rho}(\tau)) + \kappa \ddot{\alpha}^{\nu}(\tau) \ddot{\alpha}_{\nu}(\tau) \dot{\alpha}_{\mu}(\tau) \right],
\end{equation}
with $\kappa = {q^2 \over 6 \pi \epsilon_0 m c^3}$, then the asymptotic solution is free of runaways \cite{barut1980eac}. 

A commonly articulated philosophical problem with this is that it requires one to know all forces that act in the future on the particle. Moreover, the particle starts accelerating before any force is actually applied. This is the problem of pre-acceleration. 

However, we are more concerned with a {\it different} problem that makes Formula \eqref{plasstransform} completely unsuited to any actual computations. The point is this: if we choose $\ddot{\alpha}(0)$, to be even {\it infinitesimally} different from the value given by \eqref{plasstransform}, the solution ceases to be bounded and turns into a violent runaway. Stated another way, equation \eqref{plasstransform} specifies a co-dimension 1 surface in the space of all possible initial conditions. The behavior of solutions to equation \eqref{eqmotionparticle} is discontinuous in this space. On this surface, the solutions are bounded; away from this surface, the solutions comprise runaways. 

From a computational point of view, this makes any numerical computations with \eqref{plasstransform} impossible since we need to specify initial conditions {\it exactly} at the critical value prescribed by \eqref{plasstransform}. From a philosophical point of view, this need to `fine tune' the initial conditions to obtain a meaningful solution is rather unappealing. 

Hence, in this paper we do not adopt this point of view but look for another method to tame the runaway solutions of classical electrodynamics.

\subsection{One-Body vs Many-Body}
\label{onevsmany}
We now pause to explain a critical difference between the many-body system and the one-body system. For the one-body system there exists a simple method that often allows us to sidestep the above complications. 

This involves solving equation \eqref{eqmotionparticle} {\it backward} in time. Here,  the runaways solutions get rapidly damped, leaving behind the `physical' solution. To our knowledge this technique was first used by Plass to obtain the solution for a charged particle orbiting an infinitely massive object with the opposite charge \cite{plass1961cee}.  

However, for the many-body problem this method fails. This is because the field 
acting on
a particle at a given time depends on the positions of the other particles in the {\it past}. Stated another way, the equations of motion for a relativistic many-body system are functional differential equations. Such equations can be solved forward in time, given past data. However, in general, they do not have a unique solution backward in time.\footnote{This raises some difficulties for the method discussed in \cite{deluca2007gie}.} Hence the trick of solving the equations of motion backward in time to get rid of the runaways cannot be applied to the many-body problem. 

In this paper, we will develop a technique that is applicable to the many-body system. As we have already mentioned, we will check this technique against extant solutions of the one-body problem obtained using the trick above. 
Thus this trick serves to bench-mark our method; while the trick itself does not generalize to the many-body case, our method does. 

\subsection{Other approaches to this problem}
\label{otherapproachesubsection}
The formidable conceptual and calculational problems outlined above are sometimes wished away by appealing to quantum mechanics. This argument goes as follows. Equation \eqref{eqmotionparticle} is only an approximate equation valid at distances large compared to the Compton wavelength of the particle. At shorter distances we should use the full quantum theory. 
Hence, according to some authors, we need not take the difficulties above seriously, since the equation that they arise from is only approximate in the first place.

We do not find this point of view satisfactory. For one, we believe that the classical theory should be well defined in its own right. Second, in the concrete physical situations where we wish to use \eqref{eqmotionparticle}, it is hardly feasible to use the entire quantum theory.  

However, the observation that 
equation 
\eqref{eqmotionparticle} is valid only at distances large compared to the Compton wavelength 
suggests another possible resolution to the problem. 
This is 
to modify \eqref{eqmotionparticle} itself at distances that are short compared to the Compton wavelength of the particle. 

This approach has been adopted by many authors. One popular idea has focused on giving the particle itself a structure. For an excellent review of models of extended particles we refer the reader to \cite{pearle1982cem}. However, it is hard to construct a Lorentz invariant theory of extended particles. When this can be done (see for example \cite{rohrlich1997dcs}), one 
needs to worry about the forces that keep the extended particle stable. In fact, additional terms coming from the energy due to the internal stresses that hold the particle together, must be added to the mass of the particle to obtain covariant equations of motion. This is the famous `4/3' problem (see \cite{rohrlich1997dcs} for a review).

A second approach is to place a momentum cutoff in the theory. Since a momentum cutoff can be only be placed after Wick rotating the momenta to Euclidean momenta, it respects neither Lorentz invariance nor gauge invariance.

We discuss a third approach to this problem in Appendix \ref{othermethods}.

\subsection{Our Technique}

To summarize the discussion above, we are looking for a modification of equation \eqref{eqmotionparticle}. We would like this equation to be a natural consequence of a modified classical theory. Furthermore, we would like such a theory to satisfy four properties:
\begin{enumerate}
\item
\label{lorentzgauge}
It should be Lorentz covariant and gauge invariant.
\item
\label{regularfuture}
Given generic past data, the theory should possess regular solutions in the future. 
\item
It should reduce to the usual Maxwell theory at large distances
\item
\label{occam}
It should be simple enough to make solutions of the many-body problem with radiation damping feasible.   
\end{enumerate}

We have seen that the method of assigning the initial acceleration via equation \eqref{plasstransform} fails on count \ref{regularfuture} above while other methods of assigning a structure to the particle fail on counts \ref{lorentzgauge} or \ref{occam}. Yet other methods discussed in Appendix \ref{othermethods} are based on ad hoc approximations that are unacceptable because they modify the second term in \eqref{eqmotionparticle} without modifying the Maxwell theory. In this paper, we will propose a modification of Maxwell electrodynamics that satisfies all four conditions above. 

We modify the propagators of classical electrodynamics at small scales in {\em position space}, while retaining Lorentz covariance and gauge invariance. The effect is \textit{as if} we had a covariant model of extended particles. Since, however, it is electrodynamics which is being modified, this allows us to sidestep issues of internal stability of the particle and the equations of motion are covariant without the addition of \textit{ad hoc} terms.  Further, the simplicity of this model makes numerical calculations feasible. At long distances our theory reduces to the usual Maxwell theory.

In the analogy with extended particles, the length-scale at which electrodynamics is modified corresponds intuitively to the `size' of the extended particle. This finite size introduces a delay in the interaction of the particle with itself and so we obtain a functional differential equation even for the one-body problem with radiation damping. Although the resulting FDEs are numerically stiff, we are able to solve them using numerical techniques recently developed by Guglielmi and Hairer \cite{guglielmi2005acb,guglielmi2001iri}.

We find the following surprising result. {\em Locally}, the force of radiation
damping is well approximated by the customary third order expression.  
However, globally our FDE has dramatically different properties from the solutions of equation \eqref{eqmotionparticle}. In the appropriate regime of parameters, our theory does \textit{not} exhibit any 
runaway behavior. The first part of our result ensures that standard calculations of radiation emitted by antennae are not affected noticeably.

As already mentioned, the simplicity of this theory makes numerical solutions of the many-body problem feasible. In this paper, we solve the important test case of a radiating particle in a central potential. The particle gradually spirals in, as is expected on physical grounds. Furthermore, we are able to read off the rate of the decay of this orbit. As already stated for this one-body problem, we can obtain this solution, even in Maxwell electrodynamics, by means of the trick explained in Section \ref{onevsmany}. The two solutions are in good agreement. 

In a subsequent paper \cite{rajuforthcoming}, we hope to apply the techniques developed here
to the full relativistic two-body problem. 

\section{The Model}
\label{modelsection}
Given a 4-current density, $j_{\mu}(x)$, the potentials, $A_{\mu}$ are given by the formula
\begin{equation}
\label{greensolution}
A_{\mu}({\bf x}) = \frac{1} {2 \pi \epsilon_0 c} \int j_{\mu}({\bf y}) G({\bf x}, {\bf y}) d^4 {\bf y} + \partial_{\mu} \chi,
\end{equation}
where the (retarded) Green function is given by \cite{Jackson:1999ce} \footnote{We will use the Dirac delta function frequently in this paper. Our presentation maintains the level of rigor typically found in the physics literature; in {\it all} the calculations below, we always integrate the delta function with another well behaved function.  However, the reader interested in the issue of products and compositions with distributions is referred to \cite{raju1982pac}. }
\begin{equation}
\label{greenfunction}
G({\bf x}, {\bf y}) =  \delta(|{\bf x} - {\bf y}|^2) \theta(x^0 - y^0).~~~ {\rm (Maxwell~electrodynamics.)}
\end{equation}
$\chi$ is an arbitrary scalar function which we will henceforth ignore by choosing Lorenz gauge $\partial_{\mu} A^{\mu} = 0$. We emphasize that this is purely for convenience; our results below are independent of this gauge choice. 
In taking the norm, we use the metric $diag(-c^2, 1,1,1)$ i.e. 
\begin{equation}
|{\bf x}|^2 = {\bf x}^{\mu} {\bf x}_{\mu} = -c^2 ({\bf x}^0)^2 + \sum_i ({\bf x}^i)^2
\end{equation}
Greek indices run over $0 \ldots 3$, while Latin indices run over the spatial directions $ 1 \ldots 3$. 
Along the lines of \cite{raju85}, in this paper we will modify the short distance structure of Maxwell electrodynamics by modifying $G$ to
\begin{equation}
\label{modgreenfunction}
G({\bf x}, {\bf y}) =  \delta(|{\bf x} - {\bf y}|^2 + d^2) \theta(x^0 - y^0).
\end{equation}
In the expression above, the null cone has been replaced by a hyperboloid;
this regulates the singularities of the theory and leads to a theory {\underline{analogous}} to a theory of 
extended particles with characteristic size $d$. It is curious that for $\sum_i (x^i - y^i)^2 = 0$, equation \eqref{modgreenfunction} forces $x^0 - y^0 = d$. For particles extended in space, one would expect the Green function to have support on a spatial interval at zero time. The behavior above is somewhat different and we may think of our particles as being extended in time! However, we caution the reader against taking this analogy too seriously. The mathematical content of the theory is specified by equation \eqref{modgreenfunction} and this is all we will make reference to below. 

The delta function in \eqref{modgreenfunction} is written in terms of manifestly Lorentz invariant quantities. Second, for a vector satisfying $({\bf x} - {\bf y})^2 = -d^2$, a Lorentz transformation cannot change the sign of $x^0 - y^0$. Hence, the Green function in \eqref{modgreenfunction} is Lorentz invariant. Consequently, our modified theory preserves Lorentz invariance.

Consider now a point particle of charge $q$, with worldline specified by a function ${\bf{\alpha}}^{\mu}(\tau)$ of the proper time, $\tau$, of the particle. We define the field strength,
\begin{equation}
F_{\mu \nu} = \partial_{[\mu} A_{\nu]},
\end{equation}
as usual.\footnote{Note that this definition ensures that even with the modified Green functions, our theory remains gauge invariant} Using equations \eqref{greensolution}, \eqref{modgreenfunction}, we find that the field produced by this particle at a point ${\bf x}$ is given by
\begin{equation}
\label{fieldstrength}
F_{\mu \nu}({\bf x})  = \frac{q} {4 \pi \epsilon_0 c \left({\bf \zeta} \cdot \dot{{\bf{\alpha}}}\right)^2}  \left( \ddot{{\bf{\alpha}}}_{[\mu} {{\bf \zeta}}_{\nu]}  -  { \dot{{\bf{\alpha}}}_{[\mu} {{\bf \zeta}}_{\nu]} \left(c^2 + {\bf \zeta} \cdot \ddot{{\bf{\alpha}}}\right)  \over {\bf \zeta} \cdot \dot{{\bf{\alpha}}}} \right),
\end{equation}
where $\dot{\bf{\alpha}}^{\mu} \equiv {d \alpha^{\mu} \over d \tau}$ and $\ddot{\bf{\alpha}}^{\mu} \equiv {d^2 \alpha^{\mu} \over d \tau^2}$ are evaluated at the {\em retarded proper time} $\tau_0$, which is  defined as the solution to 
\begin{equation}
\label{hyperboloid}
|{\bf x}^{\mu} - {\bf{\alpha}}^{\mu}(\tau_0)|^2 + d^2 = 0,
\end{equation}
that has ${\bf x}^0 > {\bf \alpha}^0(\tau_0)$. 
Further,  ${\bf \zeta}^{\mu} = {\bf x}^{\mu} - {\bf{\alpha}}^{\mu}(\tau_0)$. For a slow moving particle, the delay $\delta \equiv \tau - \tau_0 \sim \frac{d}{c}$. 
\begin{figure}[htb]
\caption{Self-Interaction with Modified Propagators}
\label{selfin}
\includegraphics[width=1.0\hsize]{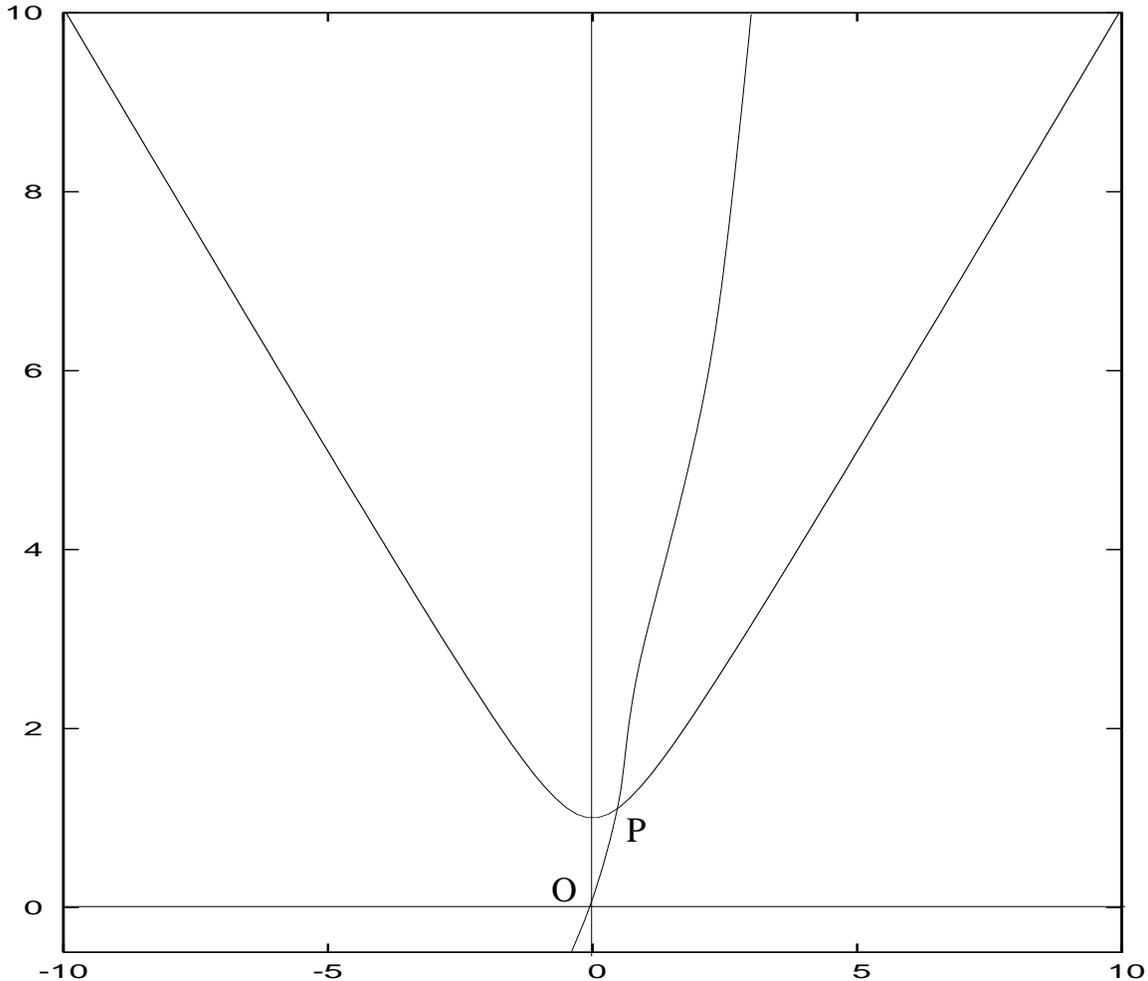}
\end{figure}
Equation \eqref{hyperboloid} defines a hyperboloid. If this hyperboloid intersects the worldline of the particle at the point `P' (see Figure \ref{selfin}), then the particle when it has reached `P' feels the effect from its past position at `O'.
The equation of motion of a particle of mass $m$ is given by \footnote{This is the usual Heaviside-Lorentz force law. Our modification of the Green function changes the fields produced by a given source but we use the same force law as in Maxwell electrodynamics}
\begin{equation}
\label{eqmotions}
\ddot{{\bf{\alpha}}}^{\mu} = {q \over m} \dot{{\bf{\alpha}}}_{\nu} F^{\mu \nu},  
\end{equation}
where $F^{\mu \nu}$ is the net field strength
and includes not only the contribution of other particles but {\it also} a contribution from the past history of the particle itself as we explained above.
Despite this, the expression for $F_{\mu \nu}$ in \eqref{fieldstrength} is free of
divergences. Moreover, the self-interaction in \eqref{eqmotions} leads to a finite
radiative damping term.  

Second, notice that even for a free particle the equation of motion \eqref{eqmotions} is in general a functional differential equation  because it relates 
$\ddot{{\bf{\alpha}}}^{\mu}(\tau)$ on the left to ${\bf{\alpha}}^{\mu}(\tau_0)$ and its derivatives on the right. To solve this FDE we need the past history of the particle, apart from initial data. For FDEs where the delay is bounded above by some $\delta_0$, $\delta < \delta_0$,  the past history of the particle specified on an interval of length $\delta_0$ (with initial data specified by continuity at the endpoint of this interval) suffices to fix a solution

\subsection{An Example}
\label{examplefdeisgood}
Qualitatively, one can understand by a simple linear example how functional differential equations help in getting rid of the runaway solutions of the Lorentz-Dirac equation. Consider the equation
\begin{equation}
\label{toy}
 f'(t) = a {f(t-\epsilon) - f(t) \over \epsilon},
\end{equation}
where $a, \epsilon > 0$ are constants. If we expand this equation to first order in $\epsilon$, we find 
\begin{equation}
\label{naiveq}
(a + 1) f'(t) =  { a \epsilon \over 2} f''(t) + {\rm O}\left(\epsilon^2\right).
\end{equation}
If we discard the higher order terms, then \eqref{naiveq} has the solution 
\begin{equation}
\label{naivesolution}
f_{\rm naive} = \kappa_1 + \kappa_2 e^{2(1 + a) t \over a \epsilon},
\end{equation}
with $\kappa_1$, $\kappa_2$ being constants of integration determined by 
the initial conditions for $f$ and $f'$. Note, that for arbitrary $\kappa_2 \neq 0$,
 these solutions grow exponentially in the future.

On the other hand an exact solution to \eqref{toy} requires past data
over the entire interval $[0, \epsilon]$. Substituting, $e^{m t}$ as a test 
solution in \eqref{toy}, we find that $m$ must satisfy the characteristic equation
\begin{equation}
\label{toycharac}
m = {a \over \epsilon} (e^{- m \epsilon} - 1).
\end{equation}
It is easy to show that {\em all} solutions to \eqref{toycharac} have ${\rm Re}(m) < 0$. Hence, \eqref{toy} actually has no runaway solutions of the form \eqref{naivesolution}. The process of  expanding FDEs to obtain ODEs, by  means of a `Taylor' series,  may lead, as above, to spurious solutions; but this is the process that is used to derive the radiation damping term.
If $\epsilon$ is very small, then locally it is indeed true that $f'(t)$ is well approximated by \eqref{naiveq}; nevertheless the global properties of the solution \eqref{naivesolution} are completely incorrect!
This leads us to expect that runaway solutions may be avoided by replacing ODEs of the Lorentz-Dirac theory with FDEs, an expectation confirmed by the calculation carried out below. 

\subsection{Many Particles}
To conclude this section, we would like to remark that the formalism discussed here works for many particles as well as it does for a single particle. If we have many particles, we calculate the field due to each particle by using \eqref{fieldstrength}. The equation of motion for each particle is now given by \eqref{eqmotions}, where $F$ is the total field strength at a point, obtained by summing all the individual field strengths. 

It is notable, that equation \eqref{eqmotions} is considerably simpler than \eqref{eqmotionparticle} which applies in  Maxwell electrodynamics. To obtain \eqref{eqmotionparticle}, we need to separate the field strength into two pieces: (a) the field strength due to all other particles and (b) the field strength due to the particle itself. The second term --- when regularized --- leads to the second term in \eqref{eqmotionparticle}. It is this term that is responsible for the runaway solutions that we have discussed. Equation \eqref{eqmotions} involves no such artificial separation. The self action of the particle on itself is regular. Since it is this term that leads to runaways in Maxwell electrodynamics and since we do not find runaways in our one-particle example below, we will not find runaway solutions in a many-body system either.

\section{Numerical Solution for a Central Force}
\label{numericalsection}

We now turn to a description of how the equations of motion \eqref{eqmotions} may be solved numerically for the important test case of a charged particle moving in 
an attractive Coulomb field.

We represent the effect of this field by the field strength
\begin{equation}
\label{coulfield}
{F}^{\rm coul}_{i 0}({\bf x})  = {-q \over 4 \pi \epsilon_0} {{\bf x}_i \over r^3},~~~ F^{\rm coul}_{i j}({\bf x}) = 0 .
\end{equation}
where $r = {\bf x}^i {\bf x}_i$.
The equation of motion for a particle of charge $q$ moving in this field is now given by \eqref{eqmotions} with 
\begin{equation}
\label{sumfields}
F_{\mu \nu} = F^{\rm coul}_{\mu \nu} + F^{\rm self}_{\mu \nu}, 
\end{equation}
where $F^{\rm self}$ is calculated, as explained in the previous section, with the help of formula \eqref{fieldstrength}.
\begin{subsection}{Choice of Code}
To solve the equations \eqref{eqmotions}, we need a numerical solver for stiff functional differential equations. The equations above are stiff for two reasons. First, the time period over which we want to solve the equations is very long compared to the delay. Second, the magnitudes of the Coulomb force and the radiation reaction differ substantially. Two popular earlier codes for FDEs, {\sc retard}\cite{hairer1993sod}  and {\sc archi}\cite{paul283uga} both use the Runge-Kutta coefficients of Dormand and Prince. While these coefficients can be used to construct excellent general-purpose solvers for ODEs, they are not appropriate for stiff equations. (For an example, of how the {\sc dopri} code fails, see \cite{rajudopri}.)

In contrast, the {\sc radar} code of Guglielmi and Hairer \cite{guglielmi2005acb,guglielmi2001iri} adapts to FDEs an implicit Runge-Kutta (collocation) method based on Radau nodes. The corresponding {\sc radau} code \cite{hairer1996sod} is a long-established code which has successfully been applied to various stiff ODEs. Another technical issue is that, although we have assumed retarded propagators, the FDE \eqref{eqmotions} is classified as a `neutral' equation, in the classification of differential equations with deviating arguments \cite{elsgolts1966itd}, rather than a retarded equation, since the second derivative of ${\bf{\alpha}}^{\mu}(\tau_0)$ appears also on the right. Both {\sc archi} and {\sc radar} codes may be used also for neutral equations, although the {\sc radar} code is significantly more efficient(by a factor of about $10^3$ in our experience) in view of its better stiffness properties. The efficient solution of such FDEs requires code that can take step sizes much larger than the delay: the {\sc radar} code allows for this possibility. The successful use of the {\sc radar} code, however, requires relatively greater preliminary work to evaluate various Jacobians.

\end{subsection}

\begin{subsection}{Rewriting the Equations for Numerical Solution}
For the numerical solution it is useful to cast the equations \eqref{eqmotions} in $3+1$ form and also rewrite them as first order equations.
It is easy to see that with the introduction of an intermediate three vector $u^i$, and an external time $t$, we may rewrite the equations  as
\begin{equation}
\label{numeq}
\begin{split}
{d {\bf{\alpha}}_i \over d t} &= {u_i \over \gamma},\\
m {d u_i \over d t} &= q  F_{i 0} + q {u^j \over \gamma} F_{i j} ,
\end{split}
\end{equation}
where $\gamma$ is determined by
\begin{equation}
\gamma = \sqrt{1 + u.u/c^2} , 
\end{equation}
and $F$ is calculated using equations \eqref{fieldstrength}, \eqref{coulfield},\eqref{sumfields}. 
\end{subsection}

\begin{subsection}{Units, Initial Condition and Parameters}
The next task is to choose appropriate units. Since we wish to deal with values that are relevant to the hydrogen atom, it is convenient to choose units so that
\begin{equation}
c=300.0,~~m_e = 2.0,~~{1 \over 4 \pi \epsilon_0} = 1.5 \times 10^3,~~q = 0.1 ,
\end{equation}
where $m_e,q$ are the mass and charge of the electron. 
This corresponds to a choice of units for length, mass, charge and time as
\begin{equation*}
\begin{split}
&3.382 \times 10^{-11} {\rm m},~~~4.555 \times 10^{-31} {\rm kg}, ~~~1.602 \times 10^{-18} {\rm C},\\ &3.384 \times 10^{-17} {\rm s}.\\ 
\end{split}
\end{equation*}
For the past history, we used a circular orbit with initial velocity appropriate for the first Bohr orbit. This is given by $v_{\rm in} = 0.00730 c$. 

For computational convenience, we will take, unless otherwise specified, $d = 1.0 \times 10^{-3}$. However, as we explain below, the solution is not very sensitive to the exact value of $d$. For comparison, the classical radius of the electron is $d_0 = {1 \over 4 \pi \epsilon_0} {q^2 \over 2 m_e c^2} = 0.417 \times 10^{-4}$ in our units. 

The mass that we actually insert into equation \eqref{numeq}, is taken to be
\begin{equation}
\label{baremass}
m = m_e - {q^2 \over 8 \pi \epsilon_0 d c^2} .
\end{equation}
This is because, as explained below, the self-action term, {\em locally}
contains a part that may be thought of as `renormalizing' the mass back to the value $m_e$. 
\end{subsection}

\section{Results}
\label{resultsection}
\begin{figure}[htb]
\caption{Solution to L.D. equation}
\label{ldfigure}
\includegraphics[width=1.0\hsize, viewport=0 0 520 480, clip]{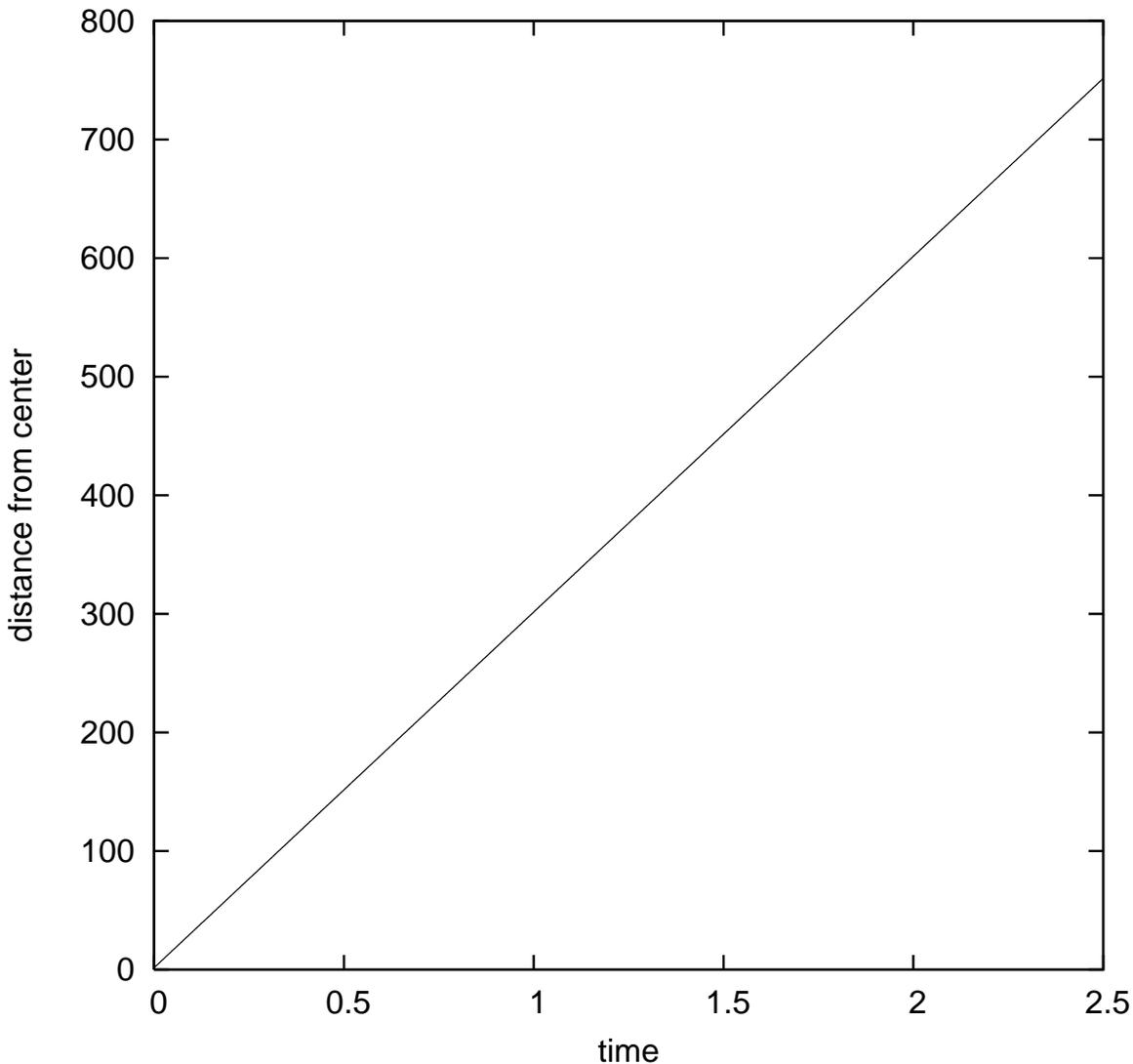}
\end{figure}

First, let us briefly investigate the solutions to the Lorentz-Dirac equation in a central force
\begin{equation}
\label{ldirac}
m {\ddot{\bf \alpha}}_{\mu} = \dot{\bf \alpha}^{\nu} F^{\rm coul}_{\mu \nu}  + {q^2 \over 6 \pi \epsilon_0 c^3} \left[\dddot{{\bf \alpha}}_{\mu} - \dot{{\bf \alpha}}_{\mu} {\ddot{{\bf \alpha}}^{\nu}\ddot{{\bf \alpha}}_{\nu} \over c^2} \right].
\end{equation}
This ODE is already stiff because of the large difference in the Coulomb term, 
the Schott term and the radiative damping term. However, it may be solved using the code {\sc radau}\cite{hairer1996sod}. The solution is shown in Figure \eqref{ldfigure}. It may be seen that the particle shoots off with the a velocity that is very close to the speed of light. This is clearly unphysical.

\begin{figure}[htb]
\caption{Solution to FDE}
\label{fdorbit}
\includegraphics[width=1.0\hsize, viewport=0 0 520 480, clip]{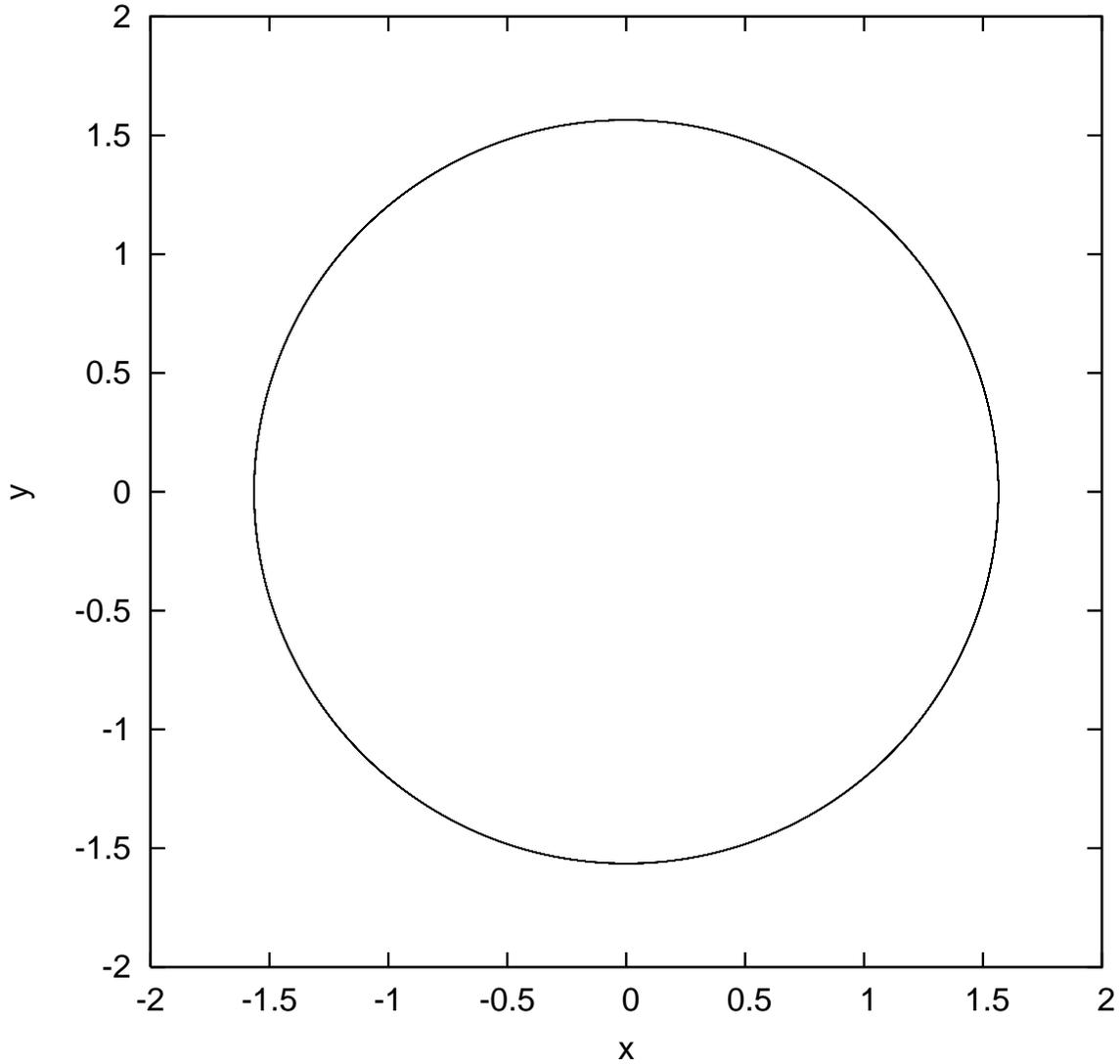}
\end{figure}

\begin{figure}[htb]
 \caption{Deviation from Circular Orbit}
\label{deviationcircular}
 \includegraphics[width=1.0\hsize, viewport=0 0 520 460, clip]{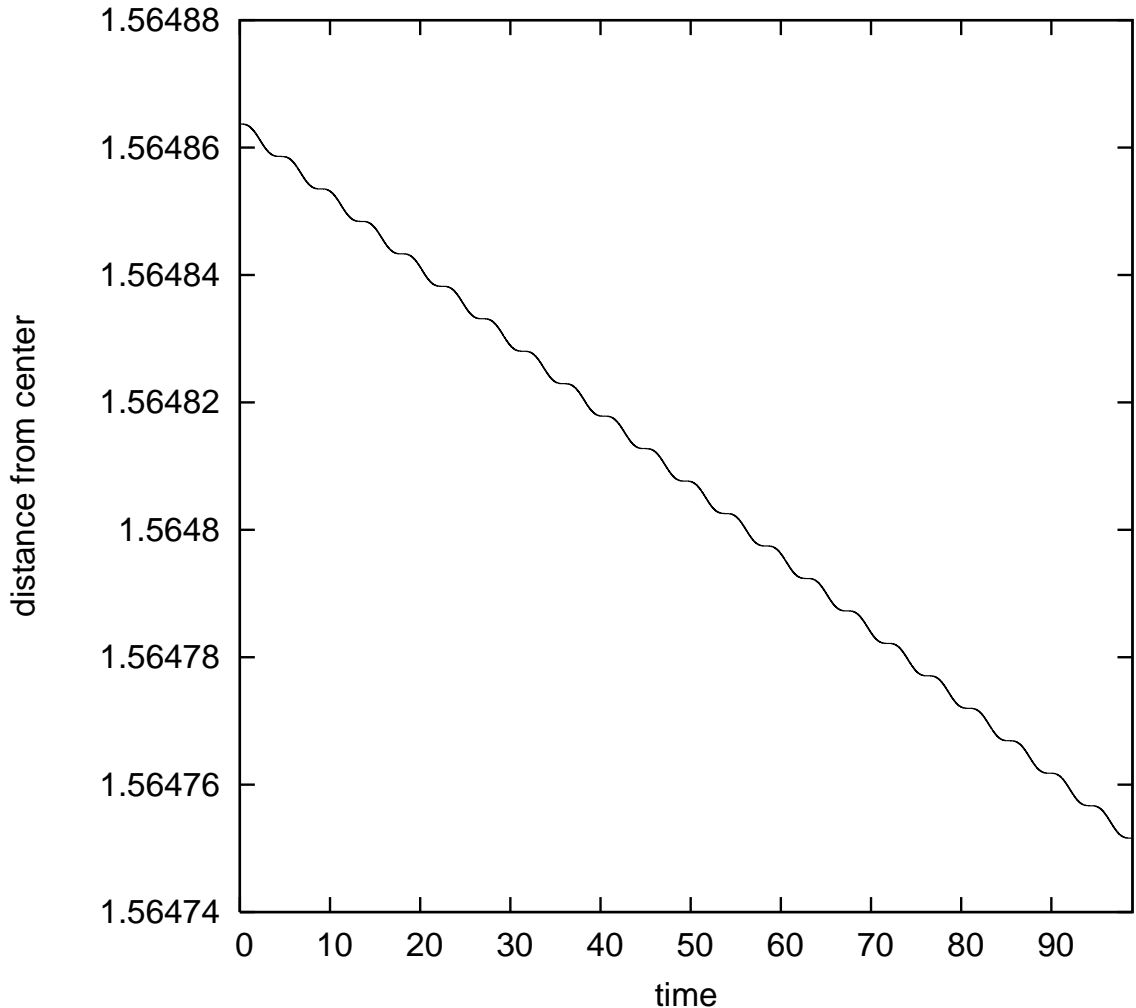}
 \end{figure}

We now turn to the solutions of the equations \eqref{numeq}.
Recall that to solve the FDE we need to specify the motion of the particle over
an appropriate past interval. As stated above, we take this past history to correspond to a circular orbit with the parameters of the first Bohr orbital.
 The solution obtained by solving \eqref{numeq} using the code {\sc radar} is then shown in Figure \ref{fdorbit}.
The damping due to radiation causes the curve above to deviate
very slightly from the circular orbit which would have been the solution
without radiation. This deviation is plotted separately in Figure \ref{deviationcircular}. 
Note, that while the particle does spiral in towards the center, this is
through a complicated elliptic spiral. The average rate of decrease
of distance towards the center may be measured from the graph in Figure \ref{deviationcircular} and is found to be 
\begin{equation}
{\delta r \over \delta t} \approx -1.13 \times 10^{-6} .
\end{equation}
This average value is in good agreement with the naive 
rate of electron decay, ${\delta r \over \delta t} \sim {-4 c \over 3 r^2} \left({q^2 \over 4 \pi \epsilon_0 m_e c^2}\right)^2$, which is derived by assuming that the orbit remains circular
while losing energy at a rate given by the Abraham Lorentz formula in each revolution 

\begin{figure}[htb]
 \caption{Difference between Solutions: $d^2=10^{-6}$ and $d^2 = 10^{-7}$}
 \label{discrepancy}
 \includegraphics[width=1.0\hsize, viewport=0 0 500 460, clip]{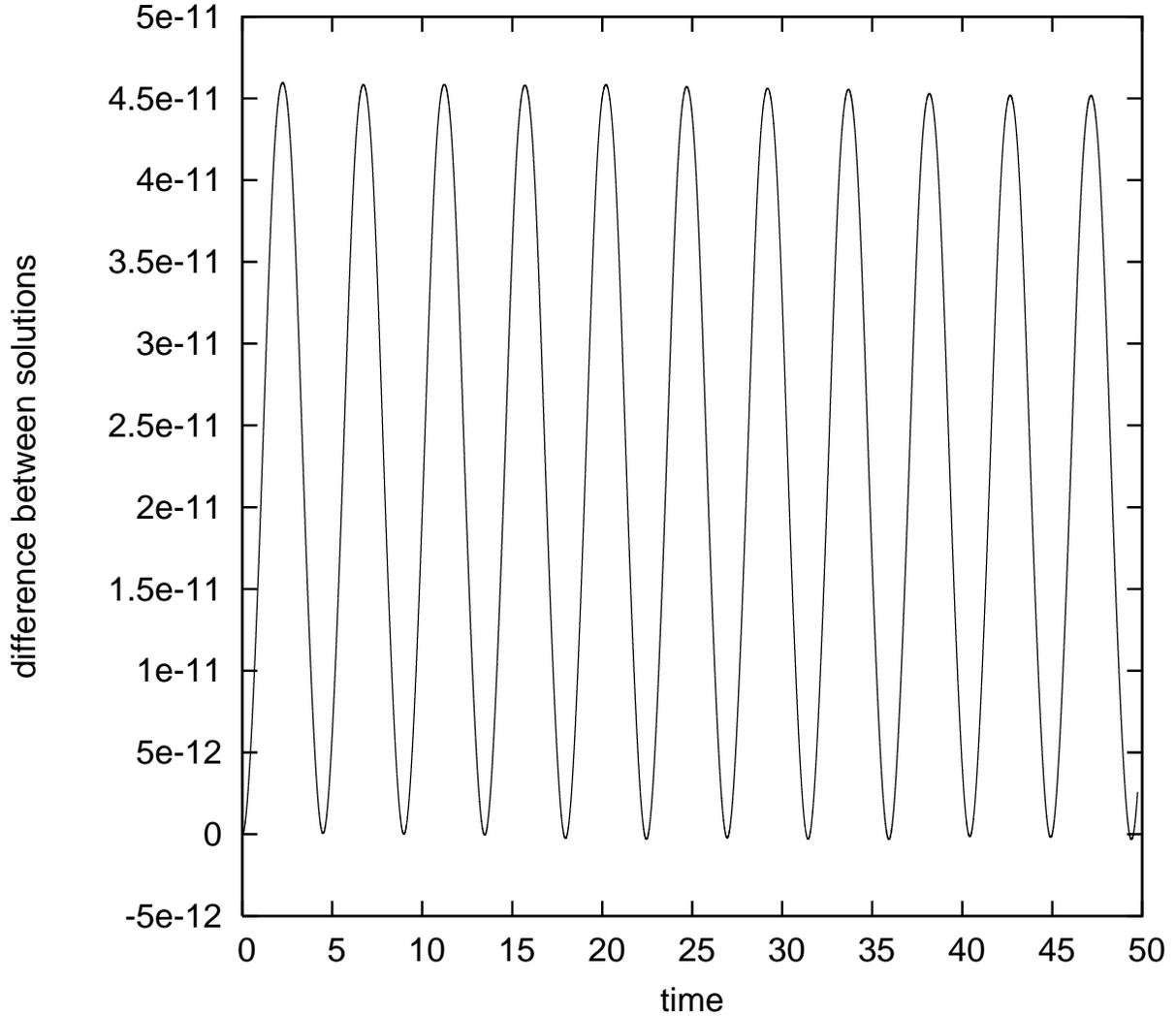}
 \end{figure}

Second, we note that the solution presented above is not very sensitive to the parameter $d$ introduced in \eqref{modgreenfunction}. In fact, if we change parameters from $d^2 = 10^{-6}$ to $d^2 =10^{-7}$, while choosing $m$ as in \eqref{baremass},  we find that the two solutions are almost identical and the 
deviation between them, in the radial coordinate, oscillates with an amplitude of $4 \times 10^{-11}$ and does not increase with time. This may be seen in Figure \ref{discrepancy}
We also note that this solution is in excellent agreement with the {\em backward} solution of the Lorentz-Dirac equation discussed above. 

Locally, the instantaneous radiation force is described very well by the 
expression:
\begin{equation}
\begin{split}
&f^{\rm rad}_i = q F^{\rm self}_{i 0} + q {u^j \over \gamma} F^{\rm self}_{i j} \approx f^{\rm emp}_i \\ &= c_1 {d \over d t} \left({d {\bf \alpha}^i \over d \tau}\right) 
+  c_2  \left({d \over d t} \left({d^2 {\bf \alpha}_i \over d \tau^2}\right) - {d {\bf \alpha}_i \over d t} {{d^2 {\bf \alpha}^{\mu} \over d \tau^2} {d^2 {\bf \alpha}_{\mu} \over d \tau^2} \over c^2} \right),
\end{split}
\end{equation}
where, to an excellent numerical approximation,
\begin{equation}
c_1 = {-q^2 \over 8 \pi \epsilon_0  d c^2}, ~~~ c_2 = {2 \over 3} \left({q^2 \over 4 \pi \epsilon_0 c^3}\right).
\end{equation}
The term $c_1$ effectively renormalizes the mass. This is the reason, we choose $m$ as in equation \eqref{baremass}.
$f^{\rm rad}_i$ oscillates as the particle moves in the orbit, and in the Table below we present, for different values of $d^2$,  the root mean square of the deviation ($f^{\rm rad} - f^{\rm emp}$), which we denote by $\sigma$, divided by the root mean square of the radiation force ($f^{\rm rad}$) itself, which we denote by$f_{\rm rms}$.  
 \begin{equation*}
 \begin{array}{c|c} d^2 & {\sigma \over f_{\rm rms}} \\ \hline  10^{-6} &  1.8 \times 10^{-10} \\  5 \times 10^{-7} & 1.2 \times 10^{-10} \\  10^{-7} & 6.8 \times 10^{-12} \\  5 \times 10^{-8}& 7.5 \times 10^{-12} \\
 \end{array}
 \end{equation*}
We see that the fit is always very good and although, initially, the fit becomes slightly better as we move to smaller $d^2$ this gain may be outweighed by 
numerical errors beyond a certain point.

The stiffness of the equations, and hence the time taken for numerical solution, increases as $d$ decreases. Eventually, the numerical solution becomes unstable for a limiting value of $d$, that we find empirically to be $d \sim 1.66 d_0$, where $d_0$ is the classical radius of the electron.

\section{Conclusions}
\begin{enumerate}
\item
In this paper we discussed the difficulties in obtaining solutions to the Lorentz-Dirac equations. We described in Section \ref{onevsmany}, why these difficulties are especially acute for the many-body problem. 
\item
Next, we proposed a modification of Maxwell electrodynamics via the modification of the Green functions in \eqref{modgreenfunction}. The modified Green function regularizes the self-action of a particle. This leads to the equation of motion \eqref{eqmotions}. A radiation damping term -- that arises from the self-action --  is automatically included in \eqref{eqmotions} since $F_{\mu \nu}$ involves a contribution from the particle itself. 
\item
{\em Instaneously}, this radiation damping term closely mimics the radiation damping term in the Lorentz-Dirac equation. However, globally, the solutions of \eqref{eqmotions} behave very different from solutions to the Lorentz-Dirac equation. We no longer find runaway solutions to the equations of motion! A heuristic understanding of how this happens is given in Section \ref{examplefdeisgood}. 
\item
We also explained that our modification of Maxwell electrodynamics was (a) Lorentz and gauge invariant (b) reduced to the usual Maxwell theory at large distances and (c) led to simple equations of motion. 
\item
As an example, we solved our new equations of motion for the important test case of a single charged particle moving in the Colulomb field defined by \eqref{coulfield}. Figure \ref{ldfigure} shows how an attempt to numerically integrate the Lorentz-Dirac equation in this physical situation leads to violent runaways. Figure \ref{fdorbit} shows the solution of our modified theory which is manifestly regular. Figure \ref{deviationcircular} shows that if one looks at this solution closely, one finds that the particle is slowly spiralling in, due to radiation losses, as one expects on physical grounds. For this simple problem, solutions to the Lorentz-Dirac equation may also be obtained using a trick explained in Section \ref{onevsmany}. The two solutions are in excellent agreement. 
\item
Having established and tested our technique in this paper, we hope to apply it to other concrete physical problems in a forthcoming paper.
\end{enumerate}

\section{Future Directions}
\label{discussionsection}
We expect that the technique we have described here will have wide applicability. An immediate  application of our technique is to the simulation of relativistic plasmas  in astrophysical contexts \cite{liang2004bal,liang2003pee}.  

Another application is to the two body problem that has seen some recent interest. The possibility that the classical solutions corresponding to quantum Bohr orbits may have distinctive features was discussed in \cite{raju2004ebp} and also conjectured from a very different viewpoint in \cite{deluca1998ehr,deluca2006stf}. These possibilities can now be investigated numerically and may lead to a better understanding of the link between differential equations with mixed type deviating arguments and quantum mechanics discussed in \cite{raju1994ttc}. (Recently, a link between pure delay differential equations and quantum mechanics was also discussed, from a somewhat different point of view, in \cite{gine2006oqm}).  

Long ago, the problem of radiative damping was taken up by Wheeler and Feynman in the hope that clarity about classical  electrodynamics might help to resolve the divergences of quantum electrodynamics. In this context, it is noteworthy that the technique of modifying the support of the propagators in position space immediately generalizes to a method of regularization in quantum field theory, in which context it was originally proposed \cite{raju85}.  
The regulator here is not merely formal, as in dimensional regularization, and moreover, unlike naive cutoffs it preserves Lorentz invariance. This may present some advantages when one is interested in studying the effect of a UV cutoff. In the quantum theory, the unitarity of the S-matrix places stringent restrictions on the propagator. This leads to additional difficulties that we do not discuss here.

Finally, in the above discussion we have taken for granted that the interactions involved must be retarded. In \cite{raju1994ttc} it was shown that the more general equations with mixed-type deviating arguments could be used to recover much of the mathematical formalism of quantum mechanics. In the present context, the use of mixed-type equations may be expected to decrease the radiative damping from the full value it has in the retarded case. However, solving mixed-type equations involves additional mathematical complexities. We defer this discussion to a future paper. 

\section*{Acknowledgments}
S.R. would like to thank the Tata Institute of Fundamental Research, Mumbai, where part of this work was completed, for its hospitality.

\section*{Appendices}
\appendix
\section{Comparison with Other Methods}
\label{othermethods}
In this Appendix, we briefly discuss another popular method, that was proposed recently by Rohrlich \cite{rohrlich2001cem} to solve the problem of runaways.\footnote{After this work was completed, the paper \cite{rajeev2008esl} appeared that uses this method to analyze the behavior of a charged particle in a Coulomb potential.}
This paper 
adopts a `perturbation theory' like approach to derive a new equation for a classical point charge.  To understand this approach, consider a non-relativistic point charge moving under an external force:
\begin{equation}
m a = F_{\rm ext} + {q^2 \over 6 \pi \epsilon_0 c^3} {d a \over d t}
\end{equation}
If we now consider the second term on the right hand side to be a small perturbation on the first term, we can rewrite this equation as
\begin{equation}
\label{approximation}
m a \approx F_{\rm ext} + {q^2 \over 6 \pi \epsilon_0 c^3 m} {d F_{\rm ext} \over d t} + \ldots 
\end{equation}
A covariant version of this equation was derived in \cite{rohrlich2001cem}. Although the paper initially claimed  that the resulting equation was exact, this was shown not to be the case in \cite{rivera2002esm}. Furthermore, since the structure of the Lorentz-Dirac equation is tightly dictated by Maxwell electrodynamics a modification of the form \eqref{approximation} cannot be performed without modifying the Maxwell theory itself. Unfortunately, no such modification was proposed in \cite{rohrlich2001cem}. Finally, when relativistic effects are included we note that, once again, triple derivatives appear on the right hand side of \eqref{approximation} which leads once again to third order equations of motion.


\section{New Field Equations}
Here we briefly discuss the new field equations that result from the modification of propagators in equation \eqref{modgreenfunction}. We stress however that this is peripheral to the program that we have presented. Given any distribution of particles, we can derive the corresponding field strengths using formula \eqref{fieldstrength}. We can then calculate the subsequent motion of all particles using \eqref{eqmotions}. In this formulation of the many-body system in terms of functional differential equations, the fields are only an intermediate crutch.  

However, for some formal purposes it may be of interest to analyze the field equations themselves. The modified field equations are easily obtained by means of a Fourier transform of $G$. Since these field equations are non-local but yet {\em linear} they are more conveniently represented in Fourier space and this is what we shall present here. 

To lighten the notation below, we shall put $c = 1$, in this Appendix. 

Our task is to find the Fourier transform of the modified Green function \eqref{modgreenfunction}.
\begin{equation}
G(p) = \int e^{-i p \cdot x} \delta(x^2 + d^2) \theta(x^0) d^4 x
\end{equation}
Formally, this integral is identical to the Fourier transform --- from momentum space to position space --- of a massive propagator. Using the standard techniques, developed for this (see, for example \cite{bogoliubov1959itq}), we find
\begin{widetext}
\begin{equation}
\label{fouriergreen}
\begin{split}
&{-i G(p) \over (2 \pi)^3} = {1 \over 4 \pi} s (p^0) \delta(p^2) + i d {\cal P}\left\{{-  \theta(-p^2) \over 8 \pi \sqrt{-p^2}}  \left[ Y_1\left(d \sqrt{-p^2}\right) - i s(p^0)  J_1\left(d \sqrt{-p^2}\right) \right] \right.  \\ &+ \left.    {\theta(p^2) \over 4 \pi^2 \sqrt{p^2}} K_1\left(m \sqrt{p^2})\right) \right\} 
\end{split}
\end{equation}
\end{widetext}
where $J$ and $Y$ are Bessel Functions of the first and second kind, $K$ is a modified Bessel function \cite{Abramowitz:1965:HMF} and $s(p^0)$ is the sign of $p^0$. All terms above should be understood to be distributions and ${\cal P}$ denotes the  principal value.  Note that the expression \eqref{fouriergreen} is Lorentz invariant. This is because, for $p^2 \leq 0$, we cannot change $s(p^0)$ by means of a Lorentz transformation; all other terms in \eqref{fouriergreen} are written in terms of explicitly Lorentz invariant quantities.

 Let us consider this expression near $p^2 = 0$. In that limit,  \eqref{fouriergreen} may be written as:
\begin{equation}
\begin{split}
{-i G(p) \over (2 \pi)^3} &= {1 \over 4 \pi} s (p^0) \delta(p^2) - {\cal P}\left[{i \over 4 \pi^2 p^2} - {i d^2 \over 8 \pi^2} \ln {d \sqrt{|p^2|} \over 2}\right] \\ &+ {\cal P}\left[ {-d^2 \over 16 \pi} s(p^0) \theta(-p^2) + O\left(\sqrt{|p^2|} \ln |p^2| \right)\right]
\end{split}
\end{equation}
In the physics literature, the first two terms above are often combined 
\begin{equation}
{1 \over 4 \pi} s (p^0) \delta(p^2) - {\cal P} \left( {i \over 4 \pi^2 p^2}\right) \equiv  {-i \over 4 \pi^2 (p^2 - i \epsilon p^0)}
\end{equation}
This also tells us that, in the limit $d \rightarrow 0$,
\begin{equation}
\label{Gasdzero}
\lim_{d \rightarrow 0} G(p) = {2 \pi \over p^2 - i \epsilon p^0}
\end{equation}
which is familiar as the usual expression for the retarded Green function in momentum space.
The new field equations are now simply
\begin{equation}
A_{\mu}(p) = G(p) j_{\mu}(p),
\end{equation}
where $A(p)$ and $j(p)$ are the Fourier transforms of the vector potential and the current density.
\begin{equation}
A_{\mu}(p) = \int A_{\mu}(x) e^{-i p x} d^4 x, ~~ j_{\mu}(p) = \int j_{\mu} (x) e^{-i p x} d^4 x
\end{equation}
We note, that in the limit that $d \rightarrow 0$, as a consequence of \eqref{Gasdzero}, these equations reduce to the equations of Maxwell electrodynamics
\begin{equation}
p^2 A(p) = j(p)
\end{equation}

\bibliography{references}

\end{document}